# SELF DIFFUSION OF YUKAWA SYSTEM IN PRESENCE OF EXTERNAL MAGNETIC FIELD

Mahmuda Begum[1a] and Nilakshi Das[1]

[1]*Plasma Physics Research Laboratory, Department of Physics, Tezpur University, Tezpur, Napaam, Assam-784028, India*

Email: mbegumtu@gmail.com

ABSTRACT

Diffusion of dust particles is one of the most significant transport processes of strongly coupled dusty plasma that reflect the nature of inter particle interaction and characterize thermodynamics of the system. In this paper the effect of magnetic field on diffusion of dust particles in strongly coupled dusty plasma is investigated by using molecular dynamics simulations. Self- diffusion coefficients of Yukawa systems are obtained for a wide range of plasma parameters and magnetic field strength using Green- Kubo expression which is based on integrated velocity autocorrelation function (VACF). It is assumed that dust particles interact with each other by modified Yukawa (i.e. screened Coulomb) potential. The study gives interesting results of dust particle diffusion in magnetized plasma.

## I.Introduction

Dusty plasma [1-5] is a mixture of electrons, ions, gas atoms and micron-sized electrically charged dust particles. The dust particles acquire a significant negative charge on the order of $10^3$-$10^5$ elementary charges and can interact electro-statically with each other .This type of plasma is ubiquitous in nature (in space, in molecular dust clouds, in planetary atmospheres) and often appears in a number of technological processes (for example, fuel burning, an industrial processing of semiconductors etc). The interaction of dust particles is well described well by the Yukawa or *Debye – Hückel* potential

$$\Phi(r_{ij}) = \frac{Q_d^2}{4\pi\varepsilon_0 r_{ij}} \exp(-\frac{r_{ij}}{\lambda_D})$$

where $Q_d$ is the particle charge, $\lambda_D$ is the screening length due to electrons and ions and $r_{ij}$ is the distance between two particles. The dynamic properties of dusty plasma is characterized by two important parameters: Coulomb coupling parameter Γ and screening constant κ defined as

$$\Gamma = \frac{Z^2 e^2}{4\pi\varepsilon_0 a K_B T_d} \qquad \kappa = \frac{a}{\lambda_D}$$

Where $a = \left(\frac{3}{4\pi n_d}\right)^{1/3}$ is the mean inter-particle distance, $T_d$ is the temperature of the dust grains, $n_d$ is the dust number density. The system is called “strongly coupled” if the coulomb coupling parameter Γ i.e., the ratio of the average inter particle potential energy to the average kinetic energy, is comparable with or greater than unity.

Study of transport properties of strongly coupled dusty plasma is an important and emerging area of research due to its applicability in various dust crystal experiments in laboratory and in space plasmas. The diffusion of dust grains [6] is one of the most important transport properties for mass or heat transfer in complex plasma and it can determine the energy losses (dissipation) as well as the dynamic characteristics of systems, development of structures, phase transitions, the conditions of propagation of waves and the formation of instabilities. In early complex plasma experiments [7, 8] diffusive motion of dust particles has been observed. Self diffusion is the phenomena arising due to the collisions of dust particles among themselves. Hamaguchi et al. [9] obtained the self diffusion coefficient of Yukawa system using the Green-Kubo formula. Molecular dynamics (MD) simulation technique has been employed [9, 10] to calculate self diffusion coefficient in the fluid phase of Yukawa system in a wide range of thermo dynamical parameters.

In the past decade transport phenomena of Yukawa system have been studied both theoretically and experimentally in the absence of external magnetic field. Recently the behavior of dusty plasma under magnetic field has attracted attention [11-17] because of new experiments. The interactions [18] of electric and magnetic fields with the plasma and the also determine dust dynamical and transport properties. Recent theoretical studies showed that the magnetic field drastically alters the particle and energy transport in Yukawa system, in particular the diffusion coefficient [19] and the collective oscillation spectrum [20]. Morfill *et al.* [21] have studied dust diffusion across magnetic field lines caused by random charge fluctuations. They obtained the value of the diffusion constant by using an approximate approach, based on the geometrical consideration of the guiding center displacements due to random charge fluctuations.

The purpose of the present paper is to investigate the diffusion processes of strongly coupled 3D magnetized Yukawa system in a wide range of plasma parameters. We have performed molecular dynamics (MD) simulations to determine the velocity autocorrelation function (VACF) and the self diffusion coefficient D by using the Green-Kubo expressions. The results obtained are important from both practical and fundamental point of views.

## II. Theoretical model

We consider 3D dusty plasma of identical, spherical particles of mass $m_d$ and charge $Q_d$ immersed in a neutralizing background plasma subjected to an external magnetic field $\vec{B}$ applied along z-direction. The interaction potential around a test dust particle with a charge $Q_d$ is

$$\Phi(r) = \frac{Q_d^{\,2}}{4\pi\varepsilon_0 r_{ij} f_1} \exp\left( - {r_{ij}}/{\rho_s} \right) \tag{1}$$

where $f_1 = 1 + f$ and $f = \left(\omega_{pi}^2/\omega_{ci}^2\right), \omega_{pi}$ and $\omega_{ci}$ beings the ion plasma and ion gyro-frequencies respectively, $\rho_s = \sqrt{f_1}\lambda_{De} \equiv \left( {C_S}/{\omega_{ci}} \right)$ is the ion-acoustic gyro-radius, $\lambda_{De}$ is the electron Debye radius $C_S = \lambda_{De}\omega_{pi}$ and is the ion-acoustic speed.

The presence of magnetic field affects the shielding of dust particles and this indirectly leads to modification of the two controlling parameters Γ and κ to $\Gamma_m$ and $\kappa_m$ respectively, where

$$\Gamma_m = (\Gamma / f_1) = \frac{Q_d^{\,2}}{4\pi\varepsilon_0 a f_1} \frac{1}{K_B T_d} \tag{2}$$

$$k_m = \left( {a}/{\rho_s} \right) = \left( \frac{a}{\sqrt{f_1}\lambda_{De}} \right)$$

(3)

**Self Diffusion Coefficient:**

Velocity autocorrelation function (VACF) is one of the important dynamical characteristic of the system of particles. The self Diffusion coefficient D of a system of particles in three dimensional case can be calculated using velocity autocorrelation function (VACF) through the Green-Kubo integral formula

$$D = \frac{1}{3}\int_0^{\alpha} Z(t)\,dt \qquad (4)$$

where VACF= $Z(t)=\langle v_j(t)\cdot v_j(0)\rangle$ (5)

The brackets <…> are appearing in equation (5) represent the canonical ensemble average over all particles. The integrand Z(t) is the velocity autocorrelation function, which is calculated over all segments of the ensemble average of the velocity products at time t and an initial time $t_0$.

**III. Simulation Model**

We have developed a MD simulation code to calculate self-diffusion coefficients of strongly coupled dusty plasma. The velocity autocorrelation function (VACF) and Green-Kubo formula have been incorporated to estimate the diffusion coefficient of dust particles. The equation of motion for the $i_{th}$ dust particles may be written as

$$m_d \frac{d^2\vec{r}_i}{dt^2} = \vec{F}_i(t)$$

Where $\vec{F}_i(t) = -Q\sum\nabla_i\Phi(r_{ij}) + Q\vec{v}_i(t)\times\vec{B}$ for i= 1, 2, 3,..., N and $j \neq i$. Here, $m_d$ is mass of the dust grain, $r_i$ is the position of the grain i, $F_i$ is the force acting on the $i_{th}$ particle and $\phi_{ij}$ represents *Debye – Hückel* type of interaction potential. For our MD simulation we have taken dust grain mass $m_d = 4.0\times10^{-15}Kg$, ion mass $m_i = 1.6726\times10^{-27}Kg$, dust density $n_d = 3.74\times10^{10}m^{-3}$, ion density $n_i = 1.0\times10^{14}m^{-3}$, electron and ion charge $q_e = q_i = 1.602200\times10^{-19}C$, electron temperature $T_e = 2323.0K$ and dust grain radius $r_d = 2.0\times10^{-6}m$.The simulation is performed with 500 particles for FCC crystal structure. Periodic boundary condition is applied. The space, mass, time, velocity, energy and external magnetic field are normalized by $\lambda_D$ , $m_d$ , $\sqrt{\frac{m_d}{K_B T_d}}^{\,1}$ , $K_B T_d$ and $\sqrt{\frac{m_d \lambda_D^2}{K_B T_d}}$ .

## IV. Results and Analysis

Based on the expressions as described in section 2 and using MD simulation, values of diffusion coefficient for fluid and solid FCC state is calculated for screening parameter κ=1.4, 2.0, 3.0 and 4.5. Magnetic field is varied from 0.001T to 0.02T. We have investigated our results for different values of Coulomb coupling parameter Γstarting from gas to solid regimes

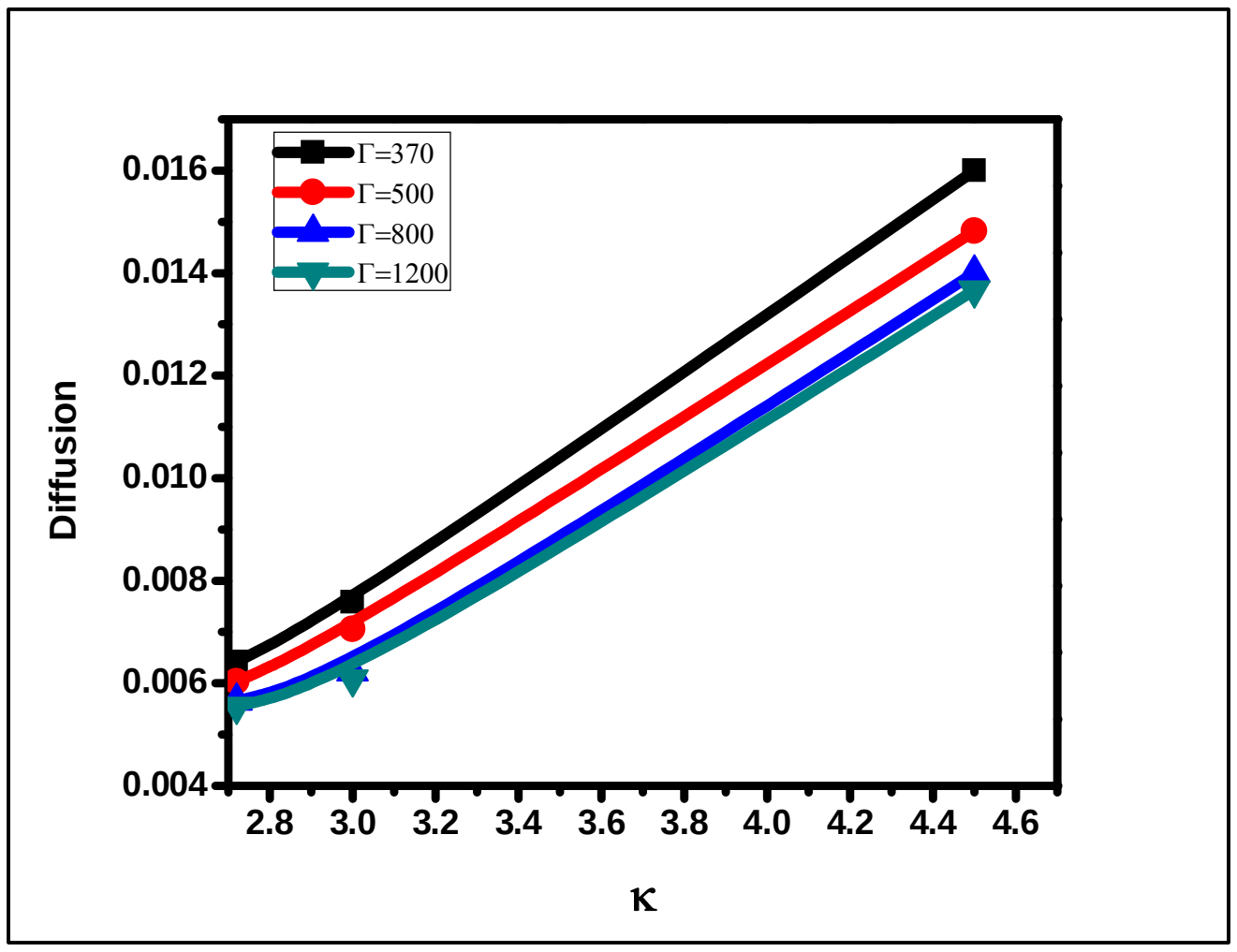

**Fig1: Diffusion coefficient vs κ plot for different values of Γ**

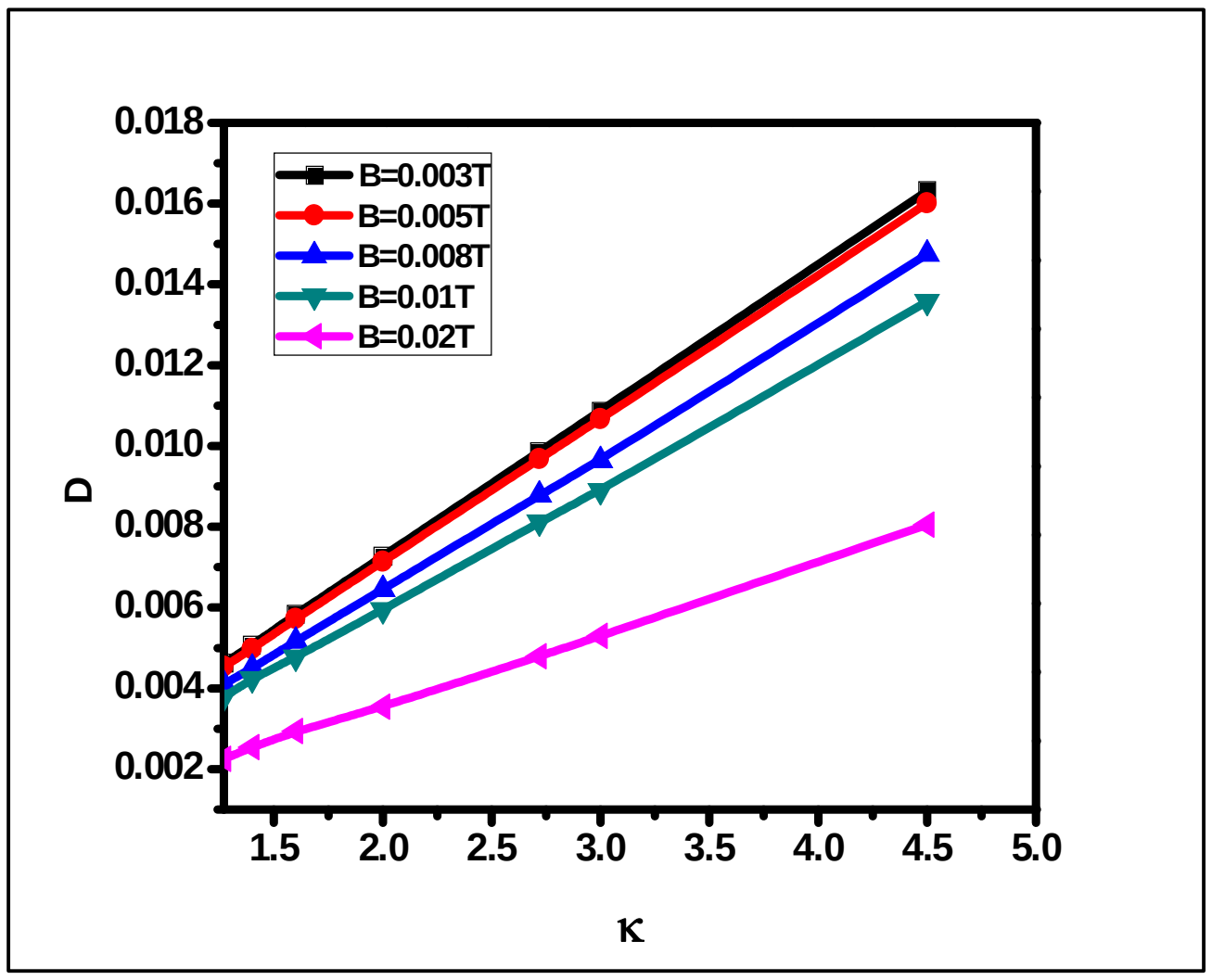

**Fig2: Diffusion coefficient vs κ plot for different values of magnetic field**

Self diffusion coefficient has been plotted in Fig1 for different values of Γ ranging from 370 to 1200. The magnetic field has been kept constant at 0.01T. Diffusion increases for higher values of κ or lower values of Γ indicating that diffusion decreases as the system goes to the strongly coupled regime.

In Fig2, diffusion coefficient has been plotted against κ for different values of magnetic field B. As the magnetic field is increased from 0.001T to 0.02T, diffusion gradually decreases. The particles are confined due to the magnetic field and these results in the reduction in diffusion.

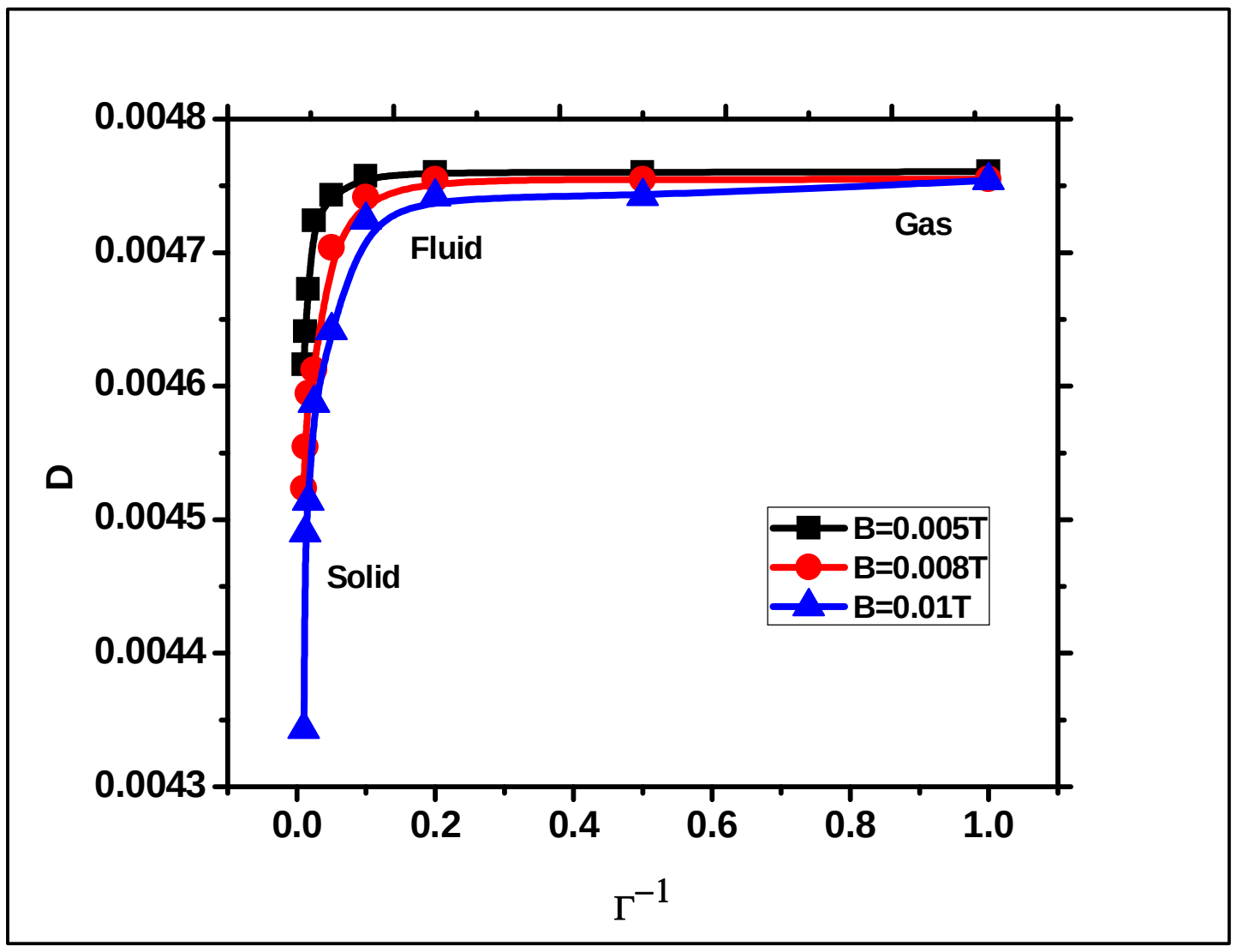

**Fig3: Diffusion coefficient vs $\Gamma^{-1}$ plot for different values of magnetic field**

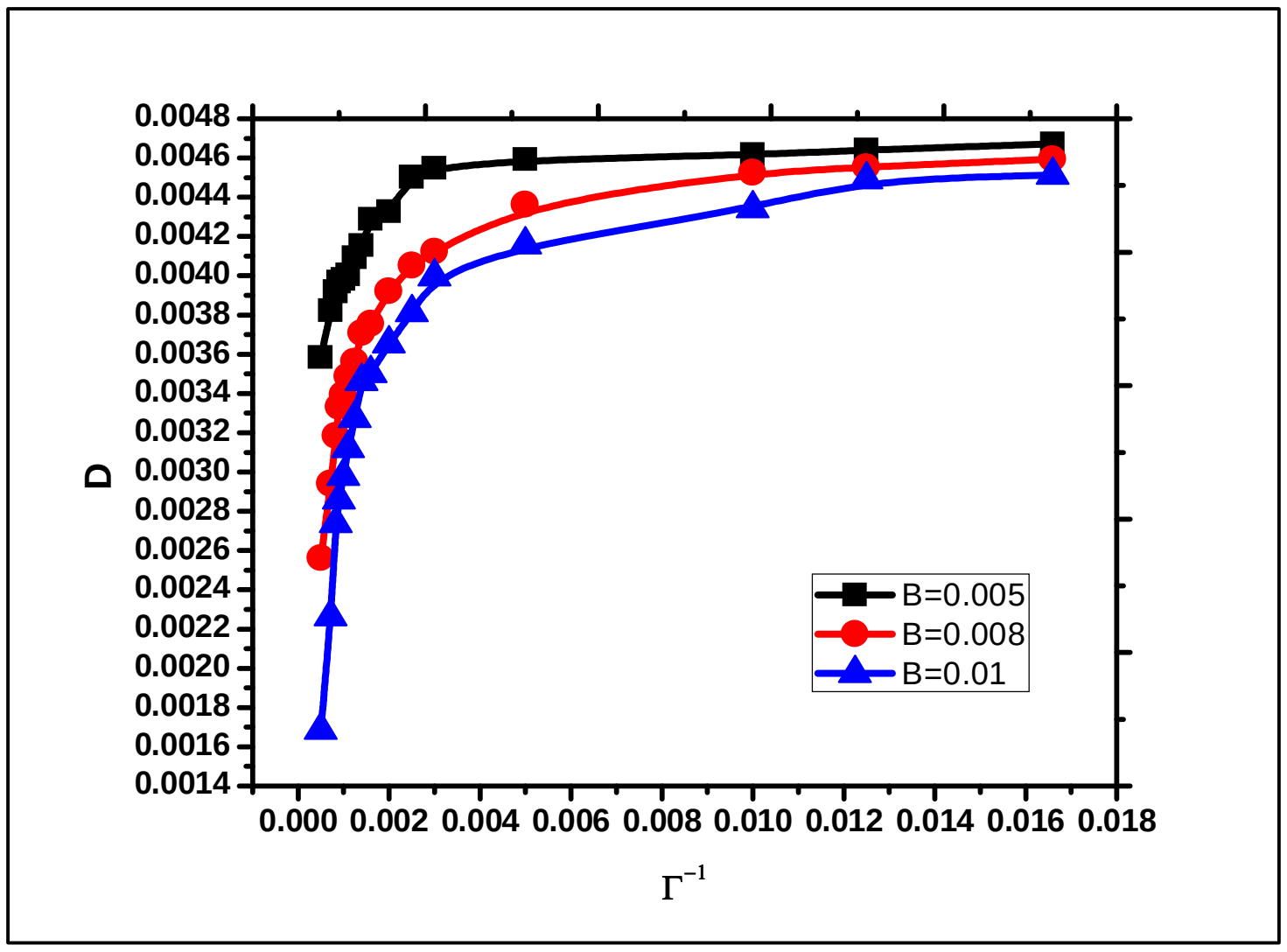

**Fig4: Diffusion coefficient vs $\Gamma^{-1}$ plot for different values of magnetic field**

In Fig3, diffusion coefficient has been plotted against $\Gamma^{-1}(T_d)$ for three different values of magnetic field. This also shows that diffusion coefficient increases with the decrease in the value of Γ. These curves gives further interesting results related to phase transition of dusty plasma. If we see the curves for a shorter range of 1/Γ (Fig4), they give idea about the transition points. At 1/Γ=0.016 ($T_d$=185), there is a change in trend of diffusion curves. For lower value of 1/Γ, the increase in diffusion is very rapid as compared to the diffusion for higher values of 1/Γ. At $T_d$=185, the lattice correlation is found to be 0.49. Around this point the transition for solid to fluid state takes place.

The diffusion coefficient has been further plotted across 1/Γ for the three different regimes- solid, liquid and gas separately. It is interesting to see how from the behavior of diffusion coefficient, one can have idea about the phase of the entire system. It is also seen that the point of transition increases for lower values of magnetic field.

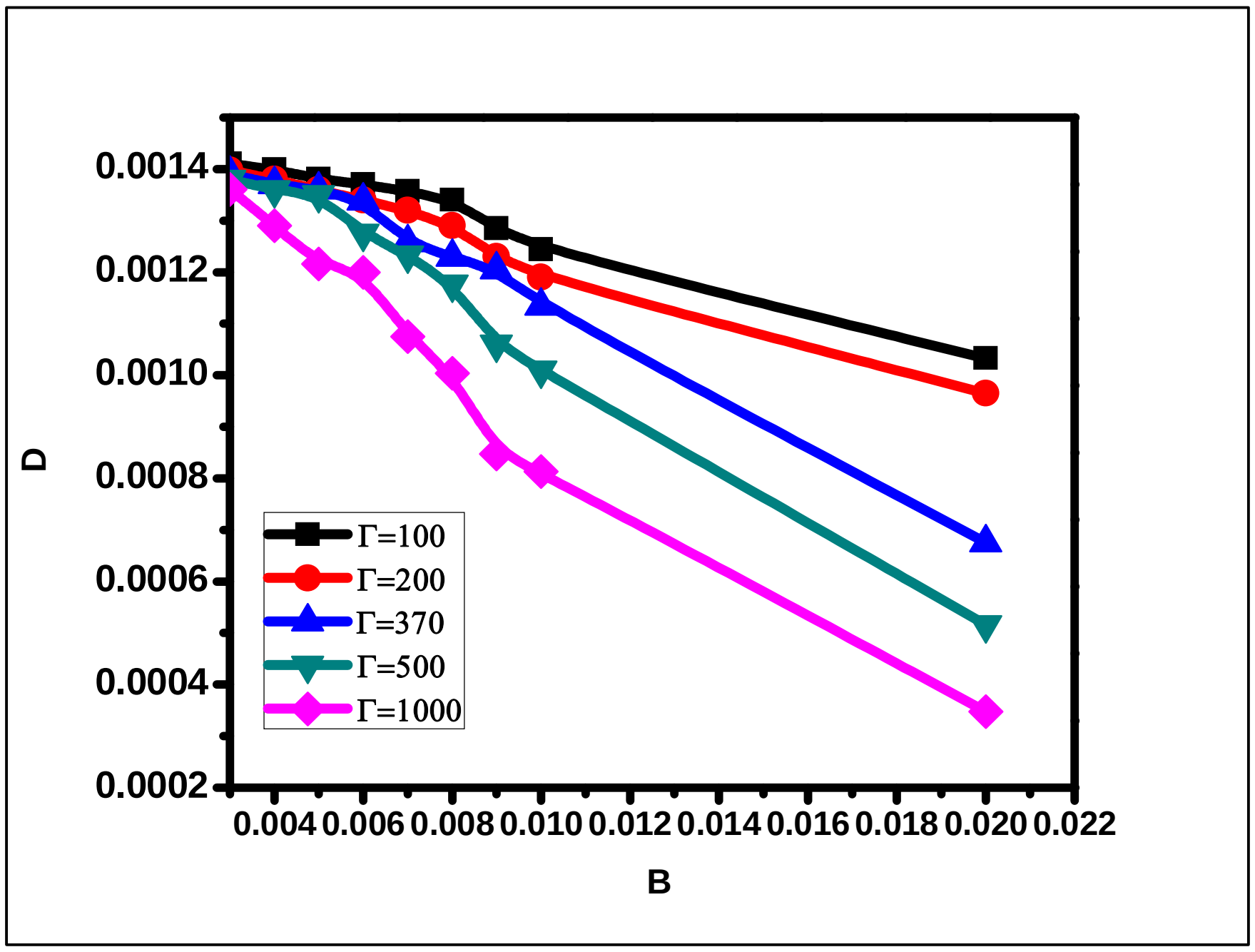

**Fig5:Diffusion coefficient vs B plot for different values of Γ**

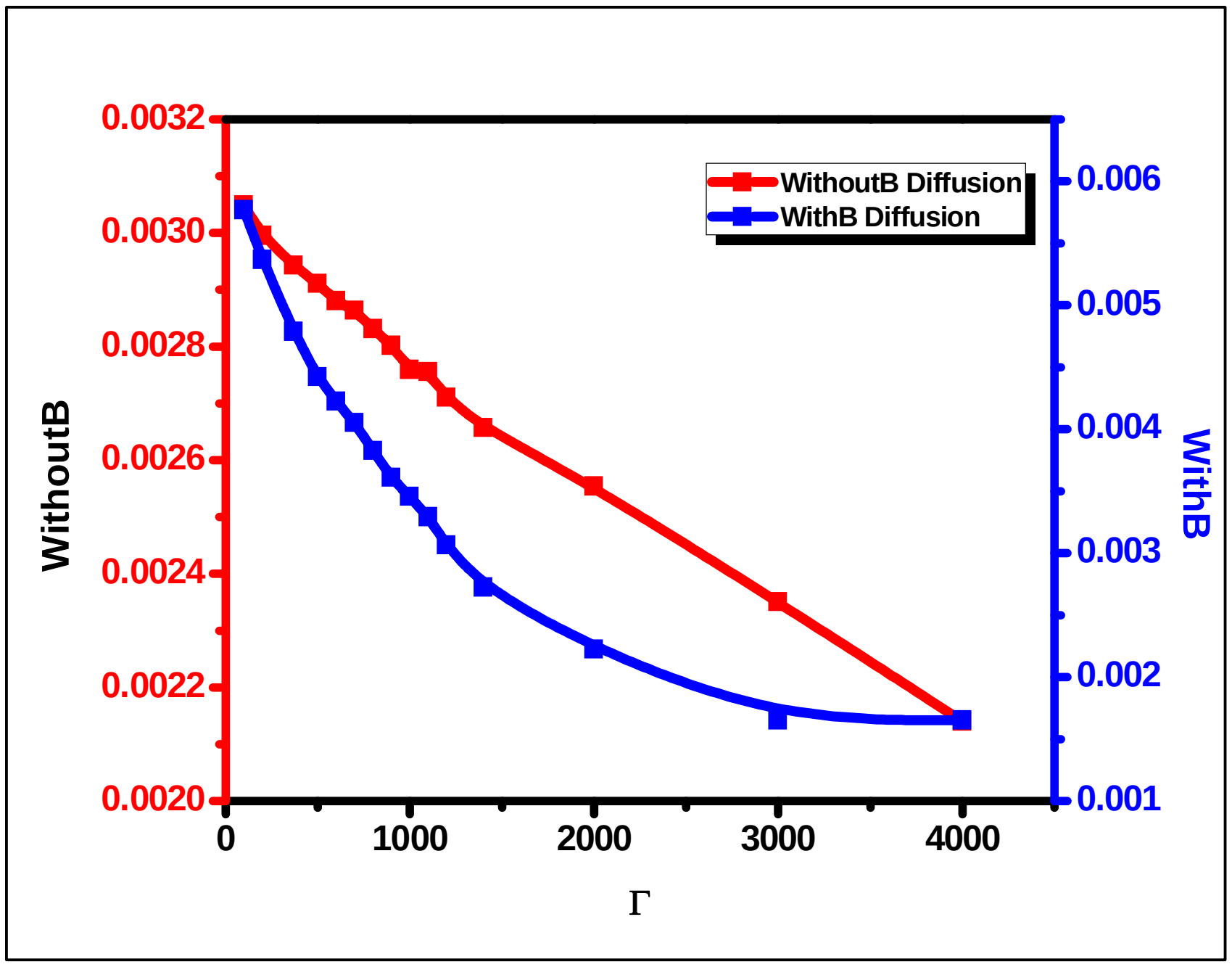

**Fig6:Diffusion coefficient vs Γ plot**

Fig5 shows the variation of diffusion coefficient with magnetic field. Here also the hump shows the point of phase transition for solid to liquid, caused by magnetic field. It is seen that the point of transition increases for higher values of coulomb coupling parameter.

A relation between diffusion coefficient and coupling parameter has been obtained from simulation data as follows:

For B=0.005T

$$D(\Gamma) = 0.00473 - 8.68238 \times 10^{-7}\Gamma + 1.28015 \times 10^{-10}\Gamma^2 + 1.05056 \times 10^{-14}\Gamma^3$$

For B=0.01T

$$D(\Gamma) = 0.00464 - 2.51427 \times 10^{-6}\Gamma + 9.20161 \times 10^{-10}\Gamma^2 - 2.0363 \times 10^{-13}\Gamma^3$$

A comparison graph of diffusion coefficient with magnetic field and without magnetic field is shown in Fig6 which indicates that the slop of the curve increases in presence of magnetic field.

## V. Summary and Conclusions

We have performed extensive molecular dynamics (MD) simulation of 3-dimensional dusty plasma in the presence of a uniform, constant magnetic field. The diffusion coefficients have been computed for a number of values of the plasma parameter and the magnetic field strength. It is observed from our study that magnetic field significantly affects the diffusion of dust particles. We have analyzed our results for different phases of the system.

## VI.The bibliography